\documentclass{aa}
\usepackage{graphics}

\begin{document}

%\thesaurus{ }
 
\title{Observational constraints on the cosmological evolution  of 
dual-population radio sources}
 
\author{M. Jamrozy\inst{1, 2}}
            
\offprints{M. Jamrozy,\\
\email: mjamrozy@astro.uni-bonn.de}

\institute{Radioastronomisches Institut der Universit\"at Bonn, Auf dem 
H\"ugel 71, D-53121 Bonn, Germany
\and Obserwatorium Astronomiczne, Uniwersytet Jagiello\'{n}ski, ul. Orla 171, 
PL-30244 Krak\'ow, Poland}
 
\date{Received  .../ Accepted ...}

\abstract{
In this paper we show normalized differential source counts n(S) at 408~MHz and 
1.4~GHz of radio sources separately for FRI and FRII classes with extended and 
compact  morphologies. The maps from the FIRST, NVSS, and WENSS surveys are 
used to define the source morphology and flux density. The counts provide a 
basis for a direct test as well as constraining the cosmological evolution of 
powerful extragalactic radio sources in terms of the dual-population model 
(Jackson \& Wall 1999), where radio sources of Fanaroff-Riley (1974) types I 
and II are regarded as two physically separate types of active galactic nuclei 
(AGN). The predicted count values are compared with the observational data to 
find the best fits for the evolution and beaming parameters, and to further 
refine the model.
} 

\authorrunning{M. Jamrozy}

\titlerunning{Dual-population radio source evolution}

\maketitle
 
\section{Introduction}
Galaxies (and/or quasars) hosting active galactic nuclei (AGN) are powerful 
extragalactic radio sources which produce jets and extended radio emitting 
regions (lobes) of plasma (Scheuer 1974). Most of these sources can be easily 
separated into two distinct classes: (i) the edge-brightened FRII sources with 
well-collimated jets, radio lobes and prominent hot spots, and (ii) 
edge-dimmed FRI sources with bending jets and peaked radio emission near the 
nucleus. There is a clear division in the luminosity between FRI and FRII, 
which lie, respectively, below and above $\rm L_{178}\approx 2\times10^{25}$ 
$\rm W Hz^{-1} sr^{-1}$ (Fanaroff \& Riley 1974). Owen \& Ledlow (1994) 
demonstrated that this separation is reproduced in the two-dimensional 
radio-optical luminosity plane in an unambiguous way. Furthermore, it is now 
quite clear that FRII and FRI sources differ radically their cosmological 
evolution (Wall 1980), i.e. the density and/or luminosity evolution rate of 
FRII is high, while that of FRI seems to be rather low.

Radio source counts n(S) involving distributions of the other observational 
parameters (i.e. luminosity, spectral index, redshift) were used to constrain 
a number of numerical models of cosmological evolution of extragalactic radio 
sources (e.g. Wall et al. 1980; Peacock \& Gull 1981; Condon 1984; Peacock 
1985; Dunlop \& Peacock 1990). All these models assumed two distinct 
populations of sources: steep-spectrum and flat-spectrum ones, which in 
principle were selected based on their spectral characteristics only. Since the 
late 70s, however,  there have been efforts to find a common scheme for some 
types of extragalactic radio sources (Rowan-Robinson 1977; Scheuer \& Readhead 
1979). These unification attempts aimed to explain the diversity of 
observational properties of extragalactic radio sources in terms of one simple 
model. The starting point for such models is the intrinsic structure of a 
radio source. Because they are not spherically symmetric, their observed form 
depends on the angle of view. The realization that orientation angle in AGN is 
of great importance for unification came with the discovery of observational 
superluminal motion in some compact radio sources (Cohen et al. 1971; Moffet 
et al. 1972) and its interpretation in terms of bulk relativistic movements of the jet's plasma 
(de Young 1972). It is obvious that when some emitting body moves in a 
relativistic manner, the radiation received by an observer is a function of 
the angle between the line of sight and the direction of motion. The small 
fraction of objects with compact morphologies are those viewed with their 
axes inclined at small angles to the line of sight. Furthermore, the beamed 
sources should be linked to an unbeamed parent population. The unified scheme 
of AGN has been quite successful in explaining differences between various 
classes of radio galaxies on the basis of the jet luminosity, its speed, 
relativistic beaming effect, orientation towards the observer, and 
environmental effects (for a review, see Urry \& Padovani 1995).

Recently,  Wall and Jackson (Wall \& Jackson 1997; Jackson \& Wall 1999) 
[hereafter WJ] introduced the {\it dual-population unified scheme}. Their 
model consists of two parent populations of powerful extragalactic radio 
sources, FRI and FRII, which appear (dependent on the viewing angle) as 
lobe-dominated (LD) steep-spectrum sources, or core-dominated (CD) 
flat-spectrum sources. 
  
Our main goal is to test the dual-population model of FRI and FRII radio 
source cosmological evolution and to refine its parameters with new 
well-defined samples of sources at two radio frequencies and within a wide 
range of radio flux density. The recent radio surveys with sufficient 
sensitivity, spatial resolution and flux-density limit, i.e. FIRST 
(Becker et al. 1995), NVSS (Condon et al. 1998)  and WENSS (Rengelink et al. 
1997), allowed us to determine morphologies, flux densities, and spectral 
indices for a large enough number of objects and thereby proved to be suitable 
for constructing such samples.

In the following section the complete source samples used in this analysis and 
the counts of extragalactic radio sources n(S) based on data at 1.4~GHz and 
408~MHz are described (Sect. 2). In Sect. 3 the application of the 
dual-population numerical model of radio source evolution is shown. In Sect. 4 
the results of modelling of the counts n(S) compared to the observational 
data and discussion are given. Finally, Sect. 5 contains the summary. 
Throughout the paper we assume $\rm H_{0}=50$ $\rm km s^{-1} Mpc^{-1}$ and 
the Einstein-de Sitter cosmological model ($\rm \Omega=\Omega_{m}=1$).

\section{Samples and observational source counts}
To construct observational source counts we selected two complete samples of 
extragalactic radio sources. The 1.4~GHz sample is selected from four 
different radio surveys, namely the FIRST and Mitchell (1983) surveys,  the 
GB/GB2 (Machalski 1998), and BDFL (Bridle et al. 1972; Bridle \& Fomalont 
1974) samples. The 408~MHz sample is based on the third Bologna radio survey 
(Ficarra et al. 1985). We exclude all starburst as well as galactic sources 
from our study. The detailed description of the 1.4~GHz and 408~MHz samples 
as well as the criterion of the source selection are described in Sect. 2.1 
and Sect. 2.2, respectively. 

We divide all sources from our two samples into four distinct types, i.e. (i) 
lobe-dominated edge-dimmed objects (LD-FRI), (ii) lobe-dominated 
edge-brightened objects (LD-FRII), (iii) compact objects with steep spectra 
(CSS), and (iv) compact objects with flat spectra (CD). A compact object is 
counted as CD-type when its low-frequency spectral index $\alpha$ 
(if $\rm S_{\nu} \propto \nu^{-\alpha}$, where $\rm S_{\nu}$ is the 
flux density at frequency $\nu$) is less than 0.6 (details about spectral 
index determinations are given below). Our CSS-class of radio sources 
includes also GigaHertz-Peaked Spectrum (GPS) sources which appear in the 
samples. In addition, in our further analysis we treat the CSS-type objects 
together with the lobe-dominated FRII ones, because it is believed that they 
are just young counterparts of the older classical doubles (de Vires et al. 
2002). Furthermore, as there is no room for lobe-dominated transient 
morphology (FRII/FRI) objects in the WJ scheme, we put all the ``hybrid'' 
sources into the LD-FRII class. In general, in this simple approach we have 
only three groups of radio sources: LD-FRI, LD-FRII  and CD sources. The 
number of sources of a particular type within the selected 1.4~GHz and 
408~MHz flux-density ranges is given in Table~1 and Table~2, respectively.

\begin{table*}[]
\begin{center}
\caption{Observational normalized differential source counts at 1.4~GHz.}
\vspace{1mm}
\begin{tabular}{cccccccccc}
& & \multicolumn{2}{c}{\bf LD-FRI }&\multicolumn{2}{c}{\bf LD-FRII}&\multicolumn{2}{c}{\bf CSS}&\multicolumn{2}{c}{\bf CD} \\                           
{\bf $\rm \Delta S$}& {\bf sky area} & {\bf n} &{\bf log(n(S))} &{\bf n} &{\bf log(n(S))} & {\bf n}& {\bf log(n(S))}&{\bf n}&{\bf log(n(S))}\\ 
{   [Jy]}       & {[sr]}&&{[$\rm sr^{-1}Jy^{1.5}$]}&&{[$\rm sr^{-1}Jy^{1.5}$]}&&{[$\rm sr^{-1}Jy^{1.5}$]}&&{[$\rm sr^{-1}Jy^{1.5}$]} \\
                &        &    &                         &     &                         &    &                        &    &                            \\
0.02$\div$0.05  & 0.0133 & 25 & $1.06^{+0.08}_{-0.10}$  &  94 &  $1.64^{+0.04}_{-0.05}$ & 71 & $1.51^{+0.05}_{-0.05}$ & 60 & $1.44^{+0.05}_{-0.06}$     \\
0.05$\div$0.10  & 0.0133 &  8 & $1.21^{+0.13}_{-0.19}$  &  47 &  $1.98^{+0.06}_{-0.07}$ & 28 & $1.75^{+0.08}_{-0.09}$ & 13 & $1.42^{+0.11}_{-0.14}$     \\
0.10$\div$0.20  & 0.0133 &  3 & $1.22^{+0.20}_{-0.37}$  &  23 &  $2.10^{+0.09}_{-0.10}$ & 19 & $2.02^{+0.09}_{-0.11}$ &  9 & $1.70^{+0.12}_{-0.18}$     \\
0.20$\div$0.25  & 0.0550 &  3 & $1.41^{+0.20}_{-0.37}$  &  16 &  $2.14^{+0.10}_{-0.12}$ &  4 & $1.54^{+0.17}_{-0.30}$ &  8 & $1.84^{+0.13}_{-0.19}$     \\
0.25$\div$0.55  & 0.0906 & 10 & $1.55^{+0.11}_{-0.15}$  &  52 &  $2.23^{+0.05}_{-0.07}$ & 19 & $1.79^{+0.09}_{-0.11}$ & 27 & $1.94^{+0.08}_{-0.09}$     \\
0.55$\div$1.00  & 0.4414 & 14 & $1.52^{+0.11}_{-0.13}$  &  93 &  $2.35^{+0.04}_{-0.05}$ & 16 & $1.58^{+0.10}_{-0.12}$ & 27 & $1.81^{+0.08}_{-0.09}$     \\
1.00$\div$2.00  & 0.4414 &  3 & $1.21^{+0.20}_{-0.38}$  &  30 &  $2.21^{+0.07}_{-0.09}$ & 11 & $1.77^{+0.11}_{-0.15}$ & 13 & $1.84^{+0.11}_{-0.14}$     \\
2.00$\div$5.00  & 4.30   & 20 & $1.44^{+0.09}_{-0.11}$  & 113 &  $2.19^{+0.04}_{-0.05}$ & 30 & $1.62^{+0.07}_{-0.09}$ & 42 & $1.76^{+0.06}_{-0.07}$     \\
5.00$\div$10.00 & 4.30   & 11 & $1.83^{+0.12}_{-0.15}$  &  12 &  $1.87^{+0.11}_{-0.15}$ &  6 & $1.57^{+0.15}_{-0.23}$ &  5 & $1.49^{+0.16}_{-0.26}$     \\
10.00$\div$20.00& 4.30   &  1 & $1.24^{+0.30}_{-0.74}$  &   3 &  $1.72^{+0.20}_{-0.37}$ &  2 & $1.54^{+0.24}_{-0.53}$ &  2 & $1.54^{+0.24}_{-0.53}$     \\
\end{tabular}
\end{center}
\end{table*}

\begin{table*}[]
\begin{center}
\caption{Observational normalized differential  source counts at 408~MHz.}
\vspace{1mm}
\begin{tabular}{cccccccccc}
& & \multicolumn{2}{c}{\bf LD-FRI }&\multicolumn{2}{c}{\bf LD-FRII}&\multicolumn{2}{c}{\bf CSS}&\multicolumn{2}{c}{\bf CD} \\                           
{\bf $\rm \Delta S$}& {\bf sky area} & {\bf n} &{\bf log(n(S))} &{\bf n} &{\bf log(n(S))} & {\bf n}& {\bf log(n(S))}&{\bf n}&{\bf log(n(S))}\\ 
{   [Jy]}      & {[sr]}&&{[$\rm sr^{-1}Jy^{1.5}$]}&&{[$\rm sr^{-1}Jy^{1.5}$]}&  &{[$\rm sr^{-1}Jy^{1.5}$]}&  &{[$\rm sr^{-1}Jy^{1.5}$]} \\
                &        &    &                         &    &                        &    &                        &    &                             \\
0.10$\div$0.20  & 0.0102 & 12 & $1.94^{+0.11}_{-0.15}$  & 62 & $2.65^{+0.05}_{-0.06}$ & 29 & $2.32^{+0.07}_{-0.09}$ & 13 & $1.97^{+0.11}_{-0.14}$      \\
0.20$\div$0.40  & 0.0327 & 10 & $1.81^{+0.12}_{-0.16}$  & 91 & $2.76^{+0.04}_{-0.05}$ & 44 & $2.45^{+0.06}_{-0.06}$ & 14 & $1.95^{+0.10}_{-0.14}$      \\
0.40$\div$0.80  & 0.0785 &  9 & $1.84^{+0.12}_{-0.18}$  & 95 & $2.86^{+0.04}_{-0.05}$ & 50 & $2.58^{+0.06}_{-0.07}$ & 16 & $2.09^{+0.09}_{-0.13}$      \\
0.80$\div$1.60  & 0.1773 &  9 & $1.94^{+0.12}_{-0.18}$  & 81 & $2.89^{+0.05}_{-0.06}$ & 31 & $2.47^{+0.07}_{-0.09}$ & 17 & $2.21^{+0.10}_{-0.12}$      \\
1.60$\div$5.00  & 0.2774 &  5 & $1.85^{+0.16}_{-0.25}$  & 58 & $2.92^{+0.05}_{-0.06}$ & 15 & $2.33^{+0.10}_{-0.13}$ & 10 & $2.15^{+0.12}_{-0.18}$      \\
5.00$\div$25.00 & 0.2774 &  2 & $2.18^{+0.23}_{-0.53}$  & 10 & $2.88^{+0.12}_{-0.17}$ &  3 & $2.35^{+0.20}_{-0.37}$ &  0 & $<1.88$                     \\
\end{tabular}
\end{center}
\end{table*}

\subsection{1.4~GHz sample}
The 1.4~GHz sample consists of objects with observed flux densities between 
20~mJy and 20~Jy (the flux scale of Baars et al. 1977 is adopted) which, 
depending on their flux, were selected from different surveys.

\noindent
{\underline{Weak source subsample.}}\\
The low flux-density subsample consists of sources with 
$\rm 20 \leq S_{1.4}\leq$ 200~mJy, selected from the FIRST and  Mitchell 
(1983) radio surveys. Both surveys were made using the Very Large Array with 
5$^{\prime\prime}$ and 15$^{\prime\prime}$  
angular resolution respectively. The flux density of each object is taken from 
the NVSS source list. To determine morphological types of the extended radio 
sources we create a contour map for each of them using the FIRST and/or NVSS 
surveys. Next we assign by eye the particular objects in the maps to FRI or 
FRII type. The completeness of the FIRST and NVSS surveys is 100\% in the 
above flux-density range, however, because of their limited baselines they 
are insensitive to extended structures (i.e. larger than 15$^{\prime}$ in the 
case of NVSS). To differentiate between compact steep-spectrum (CSS) and 
flat-spectrum (CD) sources we calculate a spectral index between 325~MHz 
(WENSS) and 1.4~GHz (NVSS). In some cases when the  source is not included in 
the WENSS catalog, its low-frequency flux density is taken from other radio 
catalogs. Because the spectral information about sources comes from the flux 
density measurements performed at two frequencies only, some of the compact
GigaHertz-Peaked Spectrum sources could be erroneously classified as CD ones. 
However, the fraction of GPS sources at 1.4~GHz constitutes at most only 
about 5\% of the entire sample (see Table 1 of Jackson \& Wall 1999).\\
The sources from the FIRST survey are selected from a small region of the sky 
between $\rm07^{h}19^{m}$ $< \alpha_{2000} <$ $\rm08^{h}00^{m}$ and 
$\rm+28^{o}45^{'}$$< \delta_{2000} <$ $\rm+31^{o}10^{'}$. The sources from the 
Mitchell survey are selected from a sky region between 
$\rm12^{h}56^{m}$ $< \alpha_{2000} <$ $\rm13^{h}20^{m}$ and 
$\rm+28^{o}16^{'}$ $< \delta_{2000} <$ $\rm+32^{o}02^{'}$. There are 400 radio 
sources in this subsample.

\noindent
{\underline{Intermediate source subsample.}}\\
This subsample is taken from the GB/GB2 revised list whose definition is 
given by Machalski (1998) and consists of 346 sources with fluxes between 
200~mJy and 2~Jy. All sources have $\rm b_{II} > 20^{o}$. The value of each 
source's spectral index and the shape of the spectrum is taken from Machalski 
(1998).

\noindent
{\underline{Strong source subsample.}}\\
The subsample of 247 sources with flux densities
$\rm2\leq S_{1.4} < 20$ Jy, found within the area of the sky defined by 
$\rm-5^{o} < \delta_{2000} < +70^{o}$, $\rm b_{II} > 20^{o}$, is taken from 
BDFL sample. The original source's flux has been transformed from the KPW 
(Kellermann et al. 1969) to the Baars scale 
($\rm S_{KPW}\times 1.029 = S_{Baars}$). In this subsample the spectral index 
is taken from the K\"uhr catalogue (K\"uhr et al. 1981).

\subsection{408~MHz sample}
The 408~MHz sample consists of 686 extragalactic sources with flux 
$\rm 0.10\leq S_{408} < $ 25~Jy selected from the B3-VLA catalogue (Vigotti 
et al. 1989). The original 408~MHz flux densities are taken from 
{\it The Third Bologna  Catalogue} (Ficarra et al. 1985). The sample includes 
objects within the R.A. range of $\rm 07^{h}03^{m}$ $\rm < \alpha_{1950} <$ 
$\rm15^{h}00^{m}$. The spectral index for sources from this sample is 
determined between 408~MHz (B3) and 1.4~GHz (NVSS).
 
\subsection{Source counts}
The observational normalized differential source counts n(S) are defined as:

\begin{equation}
\rm n(S) = \frac{nA^{-1}}{\Delta S S^{-2.5}}, 
\end{equation}

\noindent
where n is the number of sources (e.g. of lobe dominated FRI or FRII, or of 
core dominated ones) in the sky area A with flux density between $\rm S_{1}$ 
and $\rm S_{2}$ $\rm (\Delta S = S_{2}-S_{1}$), and 

\begin{equation}
\rm S=\exp\{(\log(S_{1})+\frac{\log(S_{2})-\log(S_{1})}{2}) \times \ln(10)\}.
\end{equation}

\noindent
The errors $\rm \Delta n(S)$ are derived as:

\[\rm \Delta n(S) = |\frac{(n\pm \sqrt{n}) A^{-1}}{\Delta S S^{-2.5}} - n(S)|.\]

\noindent
We also determine the observational fraction F of sources of a particular 
morphology  within a given flux-density range ($\rm \Delta$S),

\[\rm F=\frac{n}{n_{\Delta S}},\]

\noindent
and $\rm (0 \leq F \leq 1)$,
where  $\rm n_{\Delta S}$ is the number of all sources in the same flux-density 
range. The error of the fraction F is derived as: 

\begin{equation}
\rm \Delta F= \frac{F(1-F)}{{\sqrt{n}}}.
\end{equation}

\noindent
The sources from the samples described in Sect.~2 are used to construct the 
separate radio source distributions of core-dominated and lobe-dominated FRI 
or FRII objects. The resulting normalized differential counts n(S) at 1.4~GHz 
and 408~MHz are presented in Table~1 and Table~2, respectively. The columns 
give the adopted flux-density intervals, the area of sky, the observed number 
of sources of particular morphology and  the relevant differential source 
counts with estimated errors. We are aware that our procedure of morphological 
classification of radio sources is not quite perfect and may occasionally 
lead to erroneous assignments. The misclassified cases can have a significant 
effect on the resulting counts when a number of objects within the given 
interval of flux density is small. Figs.~1 and 2 show the counts described 
above, while in Fig.~3,  the observed source fraction F, both at 1.4~GHz and 
408~MHz, are plotted.

\section{Dual-population model}
The dual-population paradigm for powerful extragalactic radio sources, 
proposed by WJ, involves two parent populations: FRII and FRI type sources. 
The orientation angle of the radio axis to the plane of the sky determines the 
observed characteristics and classification of the objects as lobe-dominated 
steep-spectrum sources or core-dominated  flat-spectrum ones. The beaming 
model is applied to those distinct populations to model their respective 
contributions to the source counts. The evolution and beaming scheme can be 
used to predict the intensity-dependent n(S) distributions for beamed and 
unbeamed counterparts of FRI and FRII  radio source populations at any 
frequency. The consecutive steps to model the n(S) counts  are:
1) constructing the local luminosity function and assuming a form for the 
evolution function, 2) using the beaming model to compute the flux-densities 
and spectral indices of modeled sources, 3) computing the expected source 
counts and comparing them to the observational data, and 4) varying the 
values of model parameters and repeating the steps above until the predicted 
counts fit well the observations. The above procedure is applied to FRI and 
FRII sources separately.

\subsection{Radio luminosity function and evolution function}
The radio luminosity function $\rm \rho(L, z, \nu)dL$ specifies the comoving 
number density of sources with luminosities L to L+dL measured at frequency 
$\nu$ (in the source frame) and which are at redshift z. The local luminosity 
function $\rm \rho(L, z_{0}, \nu)$ along with an evolution function E(L, z), 
which modifies the former one, give the radio luminosity function at any epoch 
$\rm \rho(L, z, \nu)= \rho(L, z_{0}, \nu) E(L, z)$. We adopt the form of the 
local radio luminosity function directly from WJ. The evolution function 
E(L, z) is chosen as an exponential luminosity-dependent density evolution, 
first proposed by Wall et al. (1980). The evolution function has the form:

\begin{equation}
\rm E(L,z)=\exp M(L)\tau(z),
\end{equation}

\noindent
where $\tau$ is the so-called ``look-back'' time given by    

\[\rm \tau(z)=1-(1+z)^{-1.5}.\]

\noindent
We also apply a redshift cutoff $\rm z_{c}$, so that for $\rm z>z_{c}$, $\rm E(L,z)=0$.
The evolution rate M depends on a source radio power L:

\vspace{3mm}
\noindent
$\rm M(L)=M_{max} \frac{\log L- \log L_{1} }{ \log L_{2}- \log L_{1}}$ 
 \hspace{5mm} for $\rm L_{1}\leq L \leq L_{2}$,\\
$\rm M(L)=0$ \hspace{29.5mm} for $\rm L < L_{1}$,\\
$\rm M(L)=M_{max}$ \hspace{22.5mm} for $\rm L > L_{2}$.

\subsection{Relativistic beaming}
In the beaming scenario, it is assumed that all radio sources are 
intrinsically steep-spectrum, and flat-spectrum sources are merely those in 
which the core radiation is boosted in a sufficiently strong way to dominate 
the total flux density. The beaming model predicts that the observed radio 
flux density $\rm S_{\nu}$ from a given source at a given frequency $\nu$ is 
composed of flat-spectrum emission from the core and steep-spectrum from the 
extended lobes emission, $\rm S_{\nu CD}$ and $\rm S_{\nu LD}$ respectively. 
The observed core-to-extended flux ratio R, which depends on the the 
transverse (unbeamed) core dominance parameter, line-of-sight orientation of 
the jet towards the observer and the jet's speed, is given by:

\vspace{3mm}
\noindent
$\rm R(\theta,\nu)\equiv\frac{S_{\nu CD}}{S_{\nu LD}}$

\begin{equation}
\rm =R_{c}(\nu)\{[\gamma(1-\beta\cos\theta)]^{-p}+[\gamma(1+\beta\cos\theta)]^{-p}\},
\end{equation} 

\noindent
where $\rm R_{c}(\nu)$ is the frequency-dependent intrinsic core-to-extended 
flux ratio, $\rm \theta$ ($\rm 0^{o}\leq\theta\leq90^{o}$) is the angle 
between the jet axis and line-of-sight.  $\rm \gamma$ is the Lorentz factor of 
the plasma, $\rm \gamma\equiv\frac{1}{\sqrt{1-\beta^{2}}}$. For the radio 
emission with continuously ejected plasma, $\rm p=2-\alpha_{CD}$, where 
$\rm \alpha_{CD}$ is the spectral index of core emission, for which we assume 
the value of 0.0. Following WJ, we consider that the mean spectral index is
$\rm \alpha_{LD}=0.75$, and the total flux density of a source is 
$\rm S_{\nu}=S_{\nu CD}+S_{\nu LD}=S_{\nu LD}(R(\theta,\nu)+1)$.
The flux densities of the flat-spectrum component at two frequencies 
$\rm \nu_{1}$ and $\rm \nu_{2}$ are equal 
($\rm S_{\nu_{1} CD}=S_{\nu_{2} CD}$). The parameters of 
$\rm \gamma$ and $\rm R_{c}(\nu)$ are the same for each object from a given 
population of radio sources. A source is classified either as a core-dominated 
one (CD) if $\rm \theta$ is small enough so that spectral index of the entire 
source is  $\rm \alpha_{\nu_{1}}^{\nu_{2}}\leq0.6$,  or a lobe-dominated one 
(LD) for $\rm \alpha_{\nu_{1}}^{\nu_{2}}>0.6$. This spectral index is given by 
the equation (Morisawa \& Takahara 1987):  

\begin{equation}
\rm \alpha_{\nu_{1}}^{\nu_{2}}=\frac{\log[(\frac{\nu_{1}}{\nu_{2}})^{\alpha_{LD}}\frac{R(\theta, \nu_{2})+1}
{R(\theta, \nu_{1})+1}]}{\log\frac{\nu_{1}}{\nu_{2}}}.
\label{index}
\end{equation} 

\noindent
The critical observed core-to-extended flux ratio $\rm R_{min}(\theta, \nu)$, 
dividing the sources into the CD or LD type at a frequency $\nu_{2}$ is given 
by: 

\begin{equation}
\rm R_{min}(\theta, \nu_{2})=\frac
{(\frac{\nu_{1}}{\nu_{2}})^{\alpha^{\nu_{2}}_{\nu_{1}}-\alpha_{LD}}-1}
{1-(\frac{\nu_{1}}{\nu_{2}})^{\alpha^{\nu_{2}}_{\nu_{1}}}},
\label{index0}
\end{equation}

\noindent
$\rm R_{min}(\theta, \nu)$ corresponds to the critical angle $\rm \theta$, 
which can be determined from equation:

\begin{equation}
\rm \cos(\theta)=\frac{1}{\beta} -
\frac{1}{\beta \gamma} (\frac{R_{c}(\nu)}{R_{min}})^{p^{-1}},
\label{index1}
\end{equation}

\noindent
assuming that the beaming due to the counter-jet is negligible.

\subsection{Expected n(S) distributions}    
The number of sources with flux-densities S to S+dS at frequency $\nu$ are 
given by:

\begin{equation}
\rm n(S, \nu)=\int_{z=0}^{z_{max}}\int_{R_{min}}^{R_{max}} n(S, z, \nu) 
P(R(\theta,\nu))dR dz,
\label{int1}
\end{equation} 

\noindent
where $\rm n(S,z,\nu)dS$ is the number of sources in $4\pi$ sr of the sky 
with flux-densities S to S+dS at frequency $\rm \nu$ and redshift z. P(R) is 
the distribution of $\rm R(\theta, \nu)$ and in a sample of randomly 
orientated radio sources with the same values of $\rm R_{c}(\nu)$ and 
$\rm \gamma$, $\rm P(R(\theta, \nu))$ is given by: 

\[\rm P(R(\theta, \nu))dR=d(cos\theta),\]

\noindent
so equation \ref{int1} becomes:

\begin{equation}
\rm n(S, \nu)=\int_{z=0}^{z_{max}}\int_{\theta_{min}}^{\theta_{max}} n(S, z, \nu)\sin\theta 
d\theta  dz.
\label{int2}
\end{equation}

\noindent
The values of $\rm \theta_{min}$ and $\rm \theta_{max}$ depend on the source 
morphology. For CD sources  $\rm \theta_{min}=0^{o}$, and 
$\rm \theta_{max}=90^{o}$ for LD sources. $\rm \theta_{max}$ for CD is equal 
to $\rm \theta_{min}$ for LD sources and can be derived from equation 
\ref{index1}.

\noindent
The radio luminosity function $\rm \rho(L, z, \nu)$ and the number of sources 
$\rm n(S, z, \nu)$ are related in the following way:

\begin{equation}
\rm n(S,z,\nu)dS=\rho(L(S),z,\nu)dLdV,
\label{en}
\end{equation}
where dV is a volume element and is equal to $\rm 4\pi D^{2}dr$.
 
\begin{equation}
\rm D= \frac{2c \{\Omega z +(\Omega - z)[(\Omega z + 1)^{\frac{1}{2}} -1]\}} 
{H_{0}(1+z)\Omega^{2}}
\end{equation}

\noindent
and 

\begin{equation}
\rm
dr=\frac{cdz}{H_{0}(1+z)(1+\Omega z)^{\frac{1}{2}}}.
\label{er}
\end{equation}

\noindent
Inserting $\rm L=D^{2}(1+z)^{1+\alpha}S$ and equation \ref{er} into equation 
\ref{en} yields: 

\begin{equation}
\rm n(S, z, \nu)= \frac{4\pi D^{4} c (1+z)^{\alpha}\rho(L, z, \nu)}{H_{0}(1+\Omega z)^
{\frac{1}{2}} L}.
\label{en2}
\end{equation}

\noindent
It is useful to calculate the normalized counts and for that purpose we 
multiply equation \ref{en2} by $\rm S^{\frac{5}{2}}$:

\begin{equation}
\rm n(S, z, \nu)S^{\frac{5}{2}}=\frac{4\pi c (1+z)^{-\frac{5}{2} - 
\frac{3}{2}\alpha} \rho(L, z, \nu) L^{\frac{3}{2}}}
{H_{0}(1+ \Omega z)^{\frac{1}{2}} D}.
\end{equation}

\section{Results and discussion}
Following the procedure given in the preceding section we derive the 
normalized differential counts and the source fractions. The set of parameter 
values originally proposed by WJ is called `model A' (see Table 3). It is 
necessary to modify these values to fit the observational data. The 
best-modeled fit to the source counts at 1.4~GHz and 0.408~GHz is called 
`model B' and its parameter values are given also in Table 3. The values are 
determined using statistical methods, by minimizing $\chi^{2}$ evaluated 
between the observed and modeled source distributions, both at 1.4 and 
0.408~GHz. The value of $\chi^{2}$ is evaluated as:

\[\rm \chi^{2}= \sum^{N}_{i=1}(\frac{n_{data_{i}} - n_{model_{i}}}{err_{data_{i}}})^{2},\]

\noindent
where $\rm n_{data_{i}}$ is the value of the data in the ith bin, similarly 
$\rm n_{model_{i}}$ and $\rm err_{data_{i}}$ are the model value and data 
error in the ith bin, respectively. These all are determined in the logarithm 
space and the sum is taken over the N bins.\\ 

The observed and calculated n(S) distributions for 1.4 and 0.408~GHz are 
presented in Fig.~1 and Fig.~2, respectively. Fig. 3 shows the expected and 
observed source fraction.

The best-fitted  parameters of cosmological evolution and relativistic beaming 
that we estimate are slightly different from those obtained by WJ on the basis 
of the 5~GHz data. However, it is important to note that our calculations do 
not strictly follow all the steps proposed by Jackson \& Wall (1999). For 
example, we do not demand the evolution function  E(L, z) to peak at 
$\rm z_{c}$/2. Furthermore, our approach does not involve any dependence of 
the intrinsic core-to-extended flux ratio $\rm R_{c}$ on radio power. We 
estimate that the transition between evolving and non-evolving FRII sources 
appears for luminosities 
$\rm L_{151} = 10^{24.5}$ $\rm (W Hz^{-1} sr^{-1})$, whereas objects with 
$\rm L_{151} > 10^{27.95}$ $\rm (W Hz^{-1} sr^{-1})$ show maximal evolution
with the rate of $\rm M_{max}$=12.2.\\
To fit the FRI observational data, it is necessary to introduce some 
luminosity evolution for these type of sources. However,  Urry \&  Padovani 
(1995) report that the behavior of FRI radio galaxies is consistent with no 
evolution. Also, in the WJ scheme, the cosmological evolution of FRI does not 
explicitly appear, however, there are no fundamental aspects in their model 
forbidding the evolution of FRI sources in general. 
The idea that FRI should evolve is confirmed by the recent work of Snellen 
\& Best (2001), who on inspecting Hubble Deep Field images came across two 
sources of FRI morphology. This could mean that this kind of source is 
significantly more abundant at high redshifts than at the present epoch. 
They suggest also that cosmological evolution of FRI sources could be similar 
to that of FRII (i.e. is radio luminosity dependent). We propose that the 
evolution of FRI sources is similar to that of FRII, (Sect. 3.1, eq. 4). The 
evolution rate M, which depends on luminosity, has the form:

\vspace{3mm}
\noindent
$\rm M(L)=0$ \hspace{16.8mm} for $\rm L < L_{1}$,\\
$\rm M(L)=M_{max}$ \hspace{10mm} for $\rm L \geq L_{1}.$
\vspace{3mm}

\noindent
We estimate that the small fraction of FRI with luminosities 
$\rm L_{151}\geq 10^{26.5}$ $\rm (W Hz^{-1} sr^{-1})$ evolve with cosmic time.  
This is supported by the fact that at 
$\rm L_{151}\approx 10^{26.5}$ $\rm (W Hz^{-1} sr^{-1})$ in the low-frequency 
samples the fraction of objects  with observed broad-line nuclei changes 
radically from $\sim 0.4$ at higher radio luminosities to $\sim 0.1$ at lower 
luminosities (Willott et al. 2000).\\ 
The best-fit relativistic beaming parameters ($\rm R_{c}(\nu)$ and 
$\rm \gamma$) obtained are similar to those estimated by Wall \& Jackson 
(1997). They are also in agreement with those derived from observations. The 
observed core-to-extended flux ratio of individual sources spans a wide range 
of values from $\sim10^{-5}$ for the lobe-dominated sources (Morganti et al. 
1993) to $\sim10^{3}$ for the core-dominated sources (Murphy et al. 1993). The 
observations show also that there is a large spread of Lorentz factor 
$\gamma$ values from 2 to 20 (Vermulen \& Cohen 1994), while some numerical 
simulations give $4\leq\gamma\leq40$ (Urry \& Padovani 1995). However, the 
Lorentz factor value of FRII sources should be greater from those of FRI 
(Urry \& Padovani 1995).\\
Some recent papers (e.g. Caccianiga et al. 2002) suggest the existence of 
another class of sources with compact morphologies that do not originate from 
FRI or FRII populations. Our investigation shows that there is no explicit 
need to introduce any additional population of compact powerful radio 
sources, apart from the beaming version of FRI and FRII parent populations. 
However, if the values of beaming parameters given here are overestimated,
then the model predictions for compact sources will be too small when compared 
to the data.

\begin{table}[]
\begin{center}
\caption{Values of beaming and evolution parameters  for models A and B.}
\vspace{1mm}
\begin{tabular}{lrrrr}
  
{\bf Parameter}       &\multicolumn{2}{c}{\bf FRI} &\multicolumn{2}{c}{\bf FRII}\\
                      &\multicolumn{1}{c}{\bf  A}&\multicolumn{1}{c}{\bf  B} &\multicolumn{1}{c}{\bf  A} &\multicolumn{1}{c}{\bf  B}                \\
{\bf $\rm \gamma$}          & 10.3   ($\dagger$)  & 10.3  & 20.0    ($\dagger$)  & 20.0         \\  
{\bf $\rm R_{c}(\nu)(*)$}   & 0.010  ($\dagger$)  & 0.010 & 0.004   ($\dagger$)  & 0.006        \\  
{\bf $\rm \log(L_{1} (**))$}& --                  & 26.50 & 25.44   ($\ddagger$) & 24.50        \\  
{\bf $\rm \log(L_{2} (**))$}& --                  & --    & 27.34   ($\ddagger$) & 27.95        \\  
{\bf $\rm M_{max}$}         & --                  & 5.0   & 10.93   ($\ddagger$) & 12.20        \\  
{\bf $\rm z_{max}$}         & 5.0    ($\ddagger$) & 5.0   &  5.62   ($\ddagger$) &  5.62        \\    

\end{tabular}
\end{center}
$\dagger$  the value is taken from Wall \& Jackson (1997), and\\
$\ddagger$ the value is taken from Jackson \& Wall (1999),\\
(*) $\rm \nu=5~GHz$, (**) $\rm W Hz^{-1} sr^{-1}$
\end{table}

\begin{figure*}[]
\resizebox{\hsize}{130mm}{\includegraphics{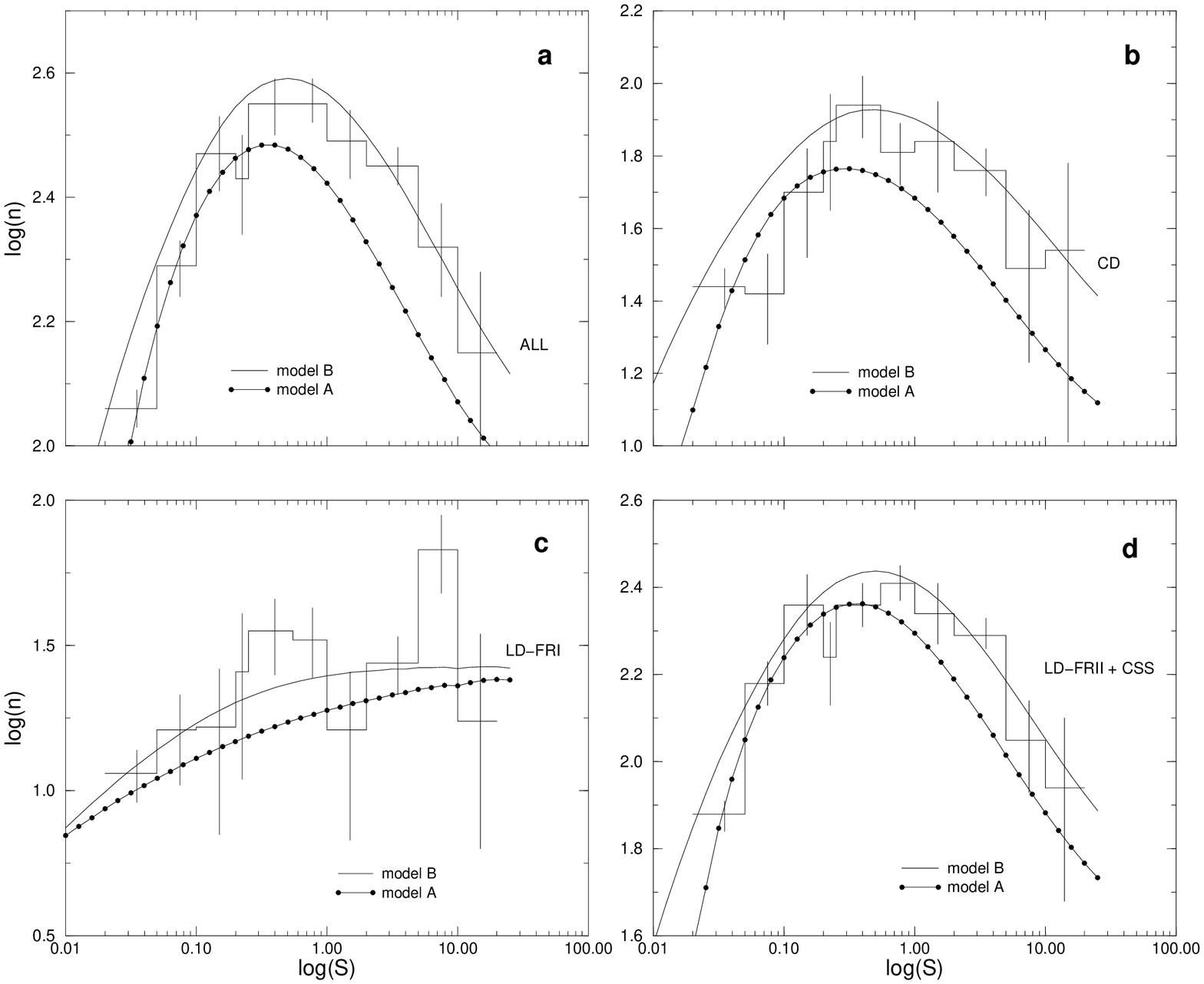}}
\caption{Normalized differential counts of sources of different morphological
types at {\bf 1.4~GHz}. Abscissa: log flux-density (Jy). Ordinate: log 
differential number of sources multiplied by $\rm S^{2.5}[sr^{-1}Jy^{1.5}]$. 
Observational data are provided in the form of stepped curve with vertical 
error bars. The solid  curve shows the counts predicted from model B and the 
dotted solid one -- counts from model A. The respective panels give the counts 
for {\bf a)} all sources in the sample, {\bf b)} compact FRI and compact FRII 
sources, {\bf c)} FRI lobe-dominated sources, and {\bf d)}  FRII 
lobe-dominated and CSS sources, (for details, see the text).}
\label{counts14}
\end{figure*} 

\begin{figure*}[]
\resizebox{\hsize}{130mm}{\includegraphics{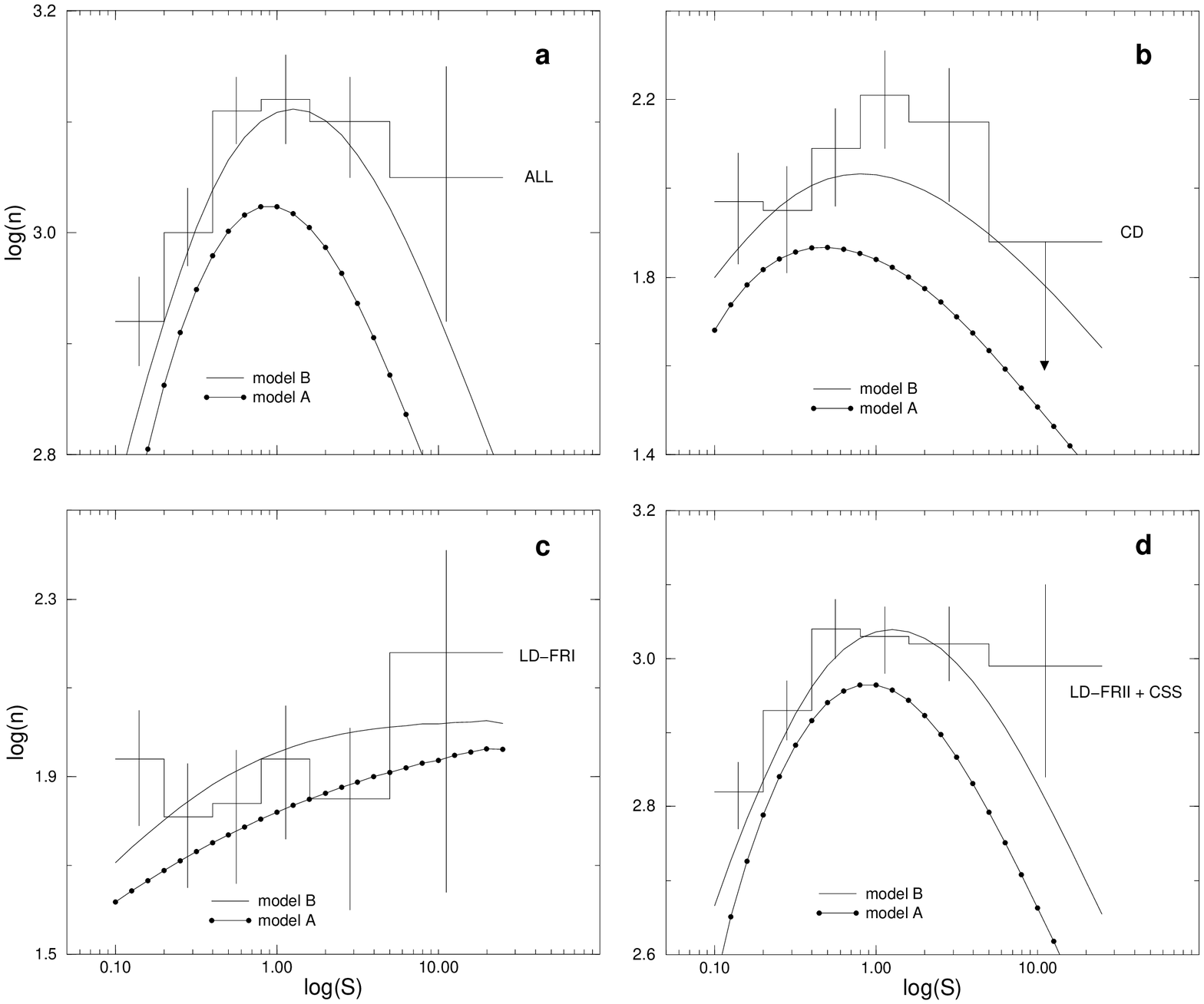}}
\caption{Normalized differential counts of sources of different morphological
types at {\bf 408~MHz}. Abscissa: log flux-density (Jy). Ordinate: log 
differential number of sources multiplied by $\rm S^{2.5}[sr^{-1}Jy^{1.5}]$. 
Observational data are provided in the form of stepped curve with vertical 
error bars. The solid  curve shows the counts predicted from model B and the 
dotted solid one -- counts from model A.
The respective panels give the counts for {\bf a)} all sources in the sample, 
{\bf b)} compact FRI and compact FRII sources, {\bf c)} FRI lobe-dominated 
sources, and {\bf d)} FRII lobe-dominated and CSS sources, (for details, see 
the text).}
\label{counts400}
\end{figure*} 
  
\begin{figure*}[]
\vspace{50mm}
\resizebox{\hsize}{80mm}{\includegraphics{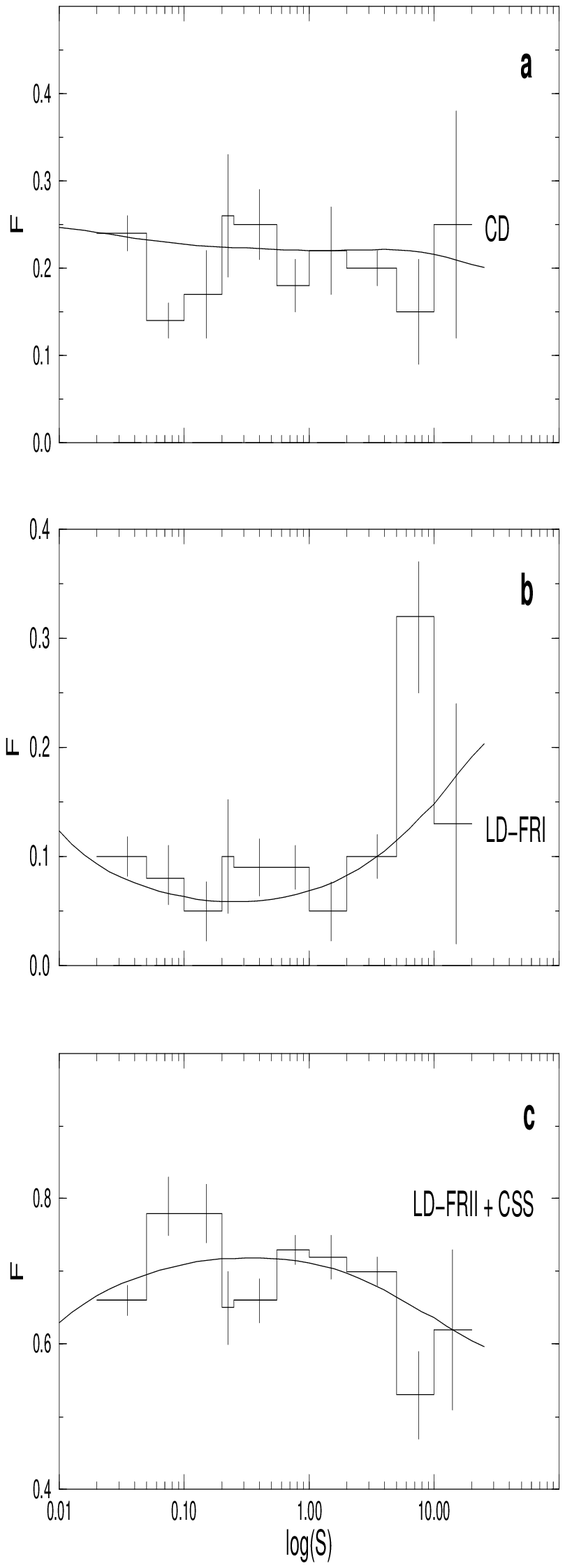} \includegraphics{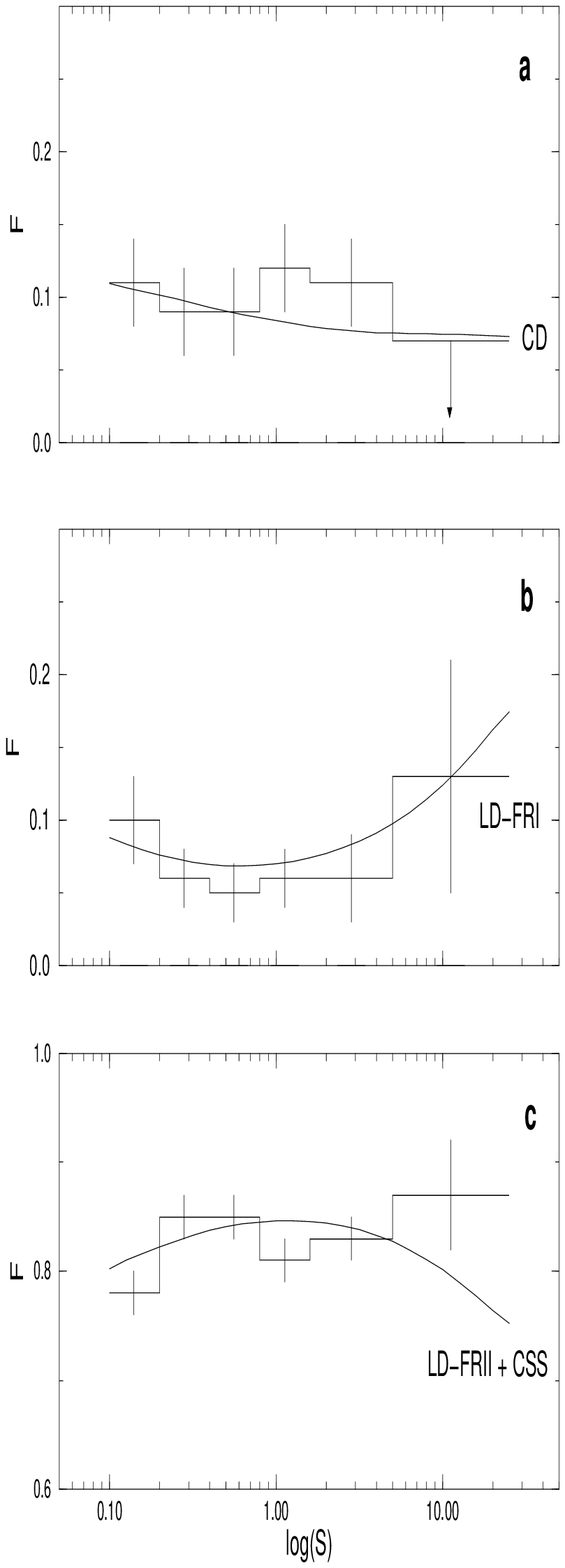}}
\caption{Fraction of {\bf a)} compact flat-spectrum sources, {\bf b)} FRI 
lobe-dominated sources, and {\bf c)} FRII  lobe-dominated together with CSS 
sources in the sample at 1.4~GHz -- {\bf left panels}, and at 408~MHz -- 
{\bf right panels}. The solid curve shows the source ratio predicted from 
model B.}
\end{figure*}

\section{Summary}
In this paper we have presented the observational log(n)-log(S) relation  
constructed for powerful extragalactic radio sources separately for different 
morphological classes (i.e. FRII and FRI lobe-, and core-dominated). The 
counts are performed over three flux-density decades (from 20~mJy to 20~Jy).
We compare the observational source distributions with the predictions of the 
dual-population WJ model. From our analysis it can be inferred that:\\
(i) the WJ dual-population evolutionary scheme enables predictions that are 
in good agreement with the observational counts of powerful extragalactic  
radio sources of a given morphological type;\\
(ii) we found a satisfactory set of physical parameter values for relativistic 
beaming and cosmological evolution which in the latter case differ slightly 
from those estimated by WJ;\\
(iii) to fit properly the observational data of FRI sources it is necessary to 
introduce an extra term of positive cosmological evolution, i.e. in the 
proposed simple approach the most luminous FRI sources evolve in a similar 
way to the FRII sources.

\begin{acknowledgements}
We are grateful to Prof.~Machalski and Dr.~Ry\'{s} for their helpful comments 
and suggestions. We also warmly thank the referee for helpful suggestions 
which have been very useful in improving and clarifying this paper. We 
acknowledge the Deutsche Forschungsgemeinschaft for the award of a 
postdoctoral felowship (GRK787).
\end{acknowledgements}

\end{document}